\documentclass[conference]{IEEEtran}

\usepackage[linesnumbered,commentsnumbered,ruled,vlined]{algorithm2e}
\usepackage{amssymb}
\usepackage{graphicx}
\usepackage{subfig}
\usepackage{makeidx}
\usepackage{multirow}
\usepackage{listings}
\usepackage[singlelinecheck=false]{caption}
\usepackage{xcolor}
\usepackage{array}
\usepackage{balance}
\usepackage{amsmath}
\usepackage{fancyvrb}
\usepackage{tabularx}
\usepackage{varwidth} 
\usepackage[inline]{enumitem}
\usepackage{lipsum}
\usepackage{enumitem}
\usepackage[T1]{fontenc}

\usepackage[referable]{threeparttablex}
\renewlist{tablenotes}{enumerate}{1}
\setlist[tablenotes]{label=\tnote{\alph*},ref=\alph*,labelindent=\tabcolsep,labelsep=.2em,leftmargin=*,align=left,before={\footnotesize}}

\graphicspath{{figure/}}

\usepackage[justification=centering]{caption}

\SetKwProg{Al}{Algorithm}{}{}
\SetAlFnt{\small\rmfamily}
\SetKw{Continue}{continue}
\SetKw{And}{and}
\SetKw{Or}{or}
\SetKw{Not}{not}
\SetKw{In}{in}
\SetKw{Null}{NULL}

\definecolor{darkspringgreen}{rgb}{0.09,0.45,0.27}

\SetCommentSty{mycommfont}

\begin{document}
	
\setcounter{secnumdepth}{3}
\newcommand{\tabincell}[2]{\begin{tabular}{@{}#1@{}}#2\end{tabular}}
\newcommand{\binCMP}{\mbox{\textsc{BinMatch}}}
\newcommand{\Bindiff}{Bindiff}
\def\tabularxcolumn#1{m{#1}}

\title{BinMatch: A Semantics-based Hybrid Approach on Binary Code Clone Analysis}

\author
{
	\IEEEauthorblockN{Yikun Hu\IEEEauthorrefmark{1}, Yuanyuan Zhang\IEEEauthorrefmark{1}, Juanru Li\IEEEauthorrefmark{1}, Hui Wang\IEEEauthorrefmark{1}, Bodong Li\IEEEauthorrefmark{1}, Dawu Gu\IEEEauthorrefmark{2}\IEEEauthorrefmark{1}}
	\IEEEauthorblockA
	{
		\IEEEauthorrefmark{1}Shanghai Jiao Tong University, Shanghai, China\\
		\IEEEauthorrefmark{2}Shanghai Institute for Advanced Communication and Data Science, Shanghai, China\\
		\{yixiaoxian, yyjess, jarod, tony-wh, uchihal, dwgu\}@sjtu.edu.cn
	}
}

\maketitle

\begin{abstract}
	
	Binary code clone analysis is an important technique which has a wide range of applications in software engineering~(e.g.,~plagiarism detection, bug detection).
	The main challenge of the topic lies in the semantics-equivalent code transformation~(e.g.,~optimization, obfuscation) which would alter representations of binary code tremendously.
	Another challenge is the trade-off between detection accuracy and coverage.
	Unfortunately, existing techniques still rely on semantics-less code features which are susceptible to the code transformation.
	Besides, they adopt merely either a static or a dynamic approach to detect binary code clones, 
	which cannot achieve high accuracy and coverage simultaneously.
	
	In this paper, we propose a semantics-based hybrid approach to detect binary clone functions.
	We execute a template binary function with its test cases, and emulate the execution of every target function for clone comparison with the runtime information migrated from that template function.
	The semantic signatures are extracted during the execution of the template function and emulation of the target function.
	Lastly, a similarity score is calculated from their signatures to measure their likeness.
	We implement the approach in a prototype system designated as \binCMP\ which analyzes \mbox{IA-32} binary code on the Linux platform.
	We evaluate \binCMP\ with eight real-world projects compiled with different compilation configurations and commonly-used obfuscation methods, totally performing over 100 million pairs of function comparison.
	The experimental results show that \binCMP\ is robust to the semantics-equivalent code transformation. Besides, it not only covers all target functions for clone analysis, but also improves the detection accuracy comparing to the state-of-the-art solutions.
\end{abstract}

\IEEEpeerreviewmaketitle

\newsavebox{\globaltemplate}
\begin{lrbox}{\globaltemplate}
	\begin{minipage}{.45\linewidth}
		\begin{lstlisting}[basicstyle=\ttfamily\scriptsize, language={[x86masm]Assembler}, numberstyle=\tiny, tabsize=4, keywordstyle=\color{blue!70}, commentstyle=\color{red!50!green!50!blue!50}, rulesepcolor=\color{red!20!green!20!blue!20}, escapechar=!, captionpos=b, numberstyle=\tiny, numbers=left, numbersep=-11pt, belowcaptionskip=-10pt]
      mov     ecx, gvar1
      test    ecx, ecx
      mov     eax, gvar2
      add     ecx, eax
		\end{lstlisting}
	\end{minipage}
\end{lrbox}

\newsavebox{\globaltarget}
\begin{lrbox}{\globaltarget}
	\begin{minipage}{.45\linewidth}
		\begin{lstlisting}[basicstyle=\ttfamily\scriptsize, language={[x86masm]Assembler}, numberstyle=\tiny, tabsize=4, keywordstyle=\color{blue!70}, commentstyle=\color{red!50!green!50!blue!50}, rulesepcolor=\color{red!20!green!20!blue!20}, escapechar=!, captionpos=b, numberstyle=\tiny, numbers=left, numbersep=-11pt, belowcaptionskip=-10pt]
      mov     ecx, gvar1!'!
      mov     ebp, gvar2!'!
      test    ebp, ebp
      add     ebp, ecx
		\end{lstlisting}
	\end{minipage}
\end{lrbox}

\newsavebox{\switchjmp}
\begin{lrbox}{\switchjmp}
	\begin{lstlisting}[basicstyle=\ttfamily\scriptsize, language={[x86masm]Assembler}, numberstyle=\tiny, tabsize=4, keywordstyle=\color{blue!70}, commentstyle=\color{red!50!green!50!blue!50}, rulesepcolor=\color{red!20!green!20!blue!20}, escapechar=!, captionpos=b, numberstyle=\tiny, numbers=left, numbersep=-11pt, belowcaptionskip=-10pt]
      mov     edx, [ebp-0E4h] ; load a local variable
      lea     eax, [edx-0Ah] ; get the index
      cmp     eax, 2Ah
      ja      loc_8052880 ; the default case
      jmp     dword ptr [eax*4+808F630h]; indirect jump
	\end{lstlisting}
\end{lrbox}

\section{Introduction}
Binary code clone analysis is a fundamental technique in software engineering. It has important applications in fields of software maintenance and security, for example, plagiarism detection~\cite{jhi2011value,zhang2012first}, patch code analysis~\cite{brumley2008automatic}, code searching~\cite{chandramohan2016bingo}, program comprehension~\cite{hu2016cross}, malware lineage inference~\cite{ming2017binsim,lindorfer2012lines,walenstein2007software}, known vulnerability detection~\cite{pewny2015cross,eschweiler2016discovre,feng2016scalable}, etc.

The main challenge that affects the accuracy of binary code clone analysis stems from the semantics-equivalent \emph{code transformation}~(\textbf{C1}), typically including link-time optimization of compilers and code obfuscation~\cite{egele2014blanket}.
The transformation modifies the representations of binary code. 
Even though two pieces of code are compiled from the same code base, the resulting binaries after the transformation would differ significantly on the syntax or structure level~(e.g., instructions, control flow graphs).
Another challenge is the \emph{trade-off} between detection accuracy and coverage~(\textbf{C2}), which corresponds to analyzing binary code in which manner, dynamic or static~\cite{wang2017memory}.
Dynamic methods procure rich semantics from code execution to ensure high accuracy, but they analyze only the executed code, leading to low coverage.
In contrast, static methods are able to cover all program components, while they rely more on syntax and structure features which lack semantics.
Additionally, static methods cannot decide the targets of indirect jumps and calls.
Thus, the analysis accuracy of static methods is relatively low.

In the literature, binary code clone analysis has drawn much attention.
However, existing techniques adopt either static method which depends on semantics-less features or dynamic method which merely cares about executed code.
For example, static methods discovRE~\cite{eschweiler2016discovre}, Genius~\cite{feng2016scalable}, and Kam1n0~\cite{ding2016kam1n0} extract features from control flow graphs, and measure similarity of binary functions basing on graph isomorphism.
\mbox{Multi-MH}~\cite{pewny2015cross} and BinGo~\cite{chandramohan2016bingo} capture behaviors of a binary function by sampling it with random values.
Since the random inputs lack semantics and are usually illegal for the function, they could hardly trigger the real semantics of a function.
For dynamic methods, Ming et al.~\cite{ming2017binsim}, Jhi et al.~\cite{jhi2011value}, and Zhang et al.~\cite{zhang2012first} perform analysis merely on executed code.
BLEX~\cite{egele2014blanket} pursues high code coverage at the cost of breaking normal execution of a binary function, distorting the semantics inferred from its collected features.
Therefore, it is necessary to propose a method which depends only on semantics and takes advantages of both static and dynamic techniques to detect binary code clones.

In this paper, we propose \binCMP, a semantics-based hybrid approach, to detect binary clone functions.
Given a template function, \binCMP\ firstly instruments and executes it with test cases to record its runtime information~(e.g.,~function argument values).
It then migrates the information to \emph{each} candidate target function and emulates the execution of the function.
During the execution and emulation, semantic signatures of the template and target functions are recorded.
Finally, \binCMP\ compares signatures of the template function and each target function to measure their similarity.
To overcome \textbf{C1} of semantics-equivalent code transformation, 
\binCMP\ only relies on semantic signatures extracted from the whole template or target function.
To address \textbf{C2} of the trade-off between accuracy and coverage, \binCMP\ adopts the hybrid method which captures semantic signatures in both static and dynamic manners.
By executing the template function, \binCMP\ captures its signature of rich semantics.
Then, it emulates every candidate target function with the runtime information of the template function to extract their signatures, which takes all target functions into consideration.

\begin{figure*}[t]
	\centering
	\includegraphics[scale=.37]{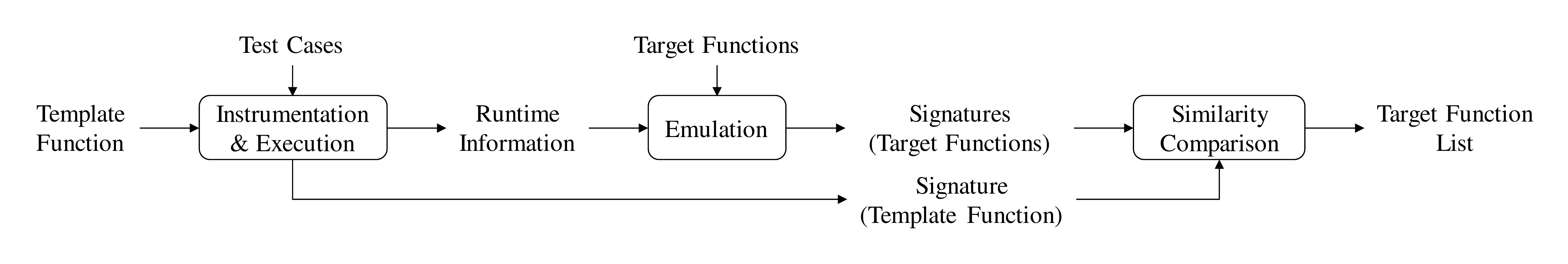}
	\caption{System Architecture of \binCMP}
	\label{fig:sys_overview}
\end{figure*}

\binCMP\ is evaluated with eight real-world projects compiled with various compilation configurations and obfuscation settings, totally performing over 100 million pairs of function comparison.
The experimental results indicate that \binCMP\ not only is robust to semantics-equivalent code transformation, but also outperforms the state-of-the-art solutions.

In summary, the contributions of this paper are as followed:
\begin{itemize}
	\item We propose a semantics-based hybrid approach to analyze binary code clones.
	The approach captures the semantic signature of a binary function in either dynamic~(execution) or static~(emulation) manner.
	Thus, it could not only detect clone functions accurately with signatures of rich semantics, but also cover all target functions under analysis.
	
	\item 
	To smooth the migration of runtime information and the emulation of a function,
	we propose novel strategies to handle global variable reading, indirect calling/jumping, and library function invocation.
	
	\item We implement the approach in a prototype system \binCMP\ which supports IA-32 binary code clone analysis on the Linux platform.
	\binCMP\ is evaluated with eight real-world projects which are compiled with different compilation configurations and obfuscation settings.
	The experimental results show that \binCMP\ is robust to the semantics-equivalent code transformation.
	Besides, it covers all candidate target functions for clone analysis, and outperforms the state-of-the-art solutions from the perspective of accuracy.
\end{itemize}

\section{Motivation and Overview}
In this section, we firstly present an example to illustrate the limitations of previous work on binary code clone analysis, which motivate our research.
Then, we explain the basic idea of our approach and show the system overview.

\subsection{Motivating Example}\label{sec:mot_ovv:mot}

It is a typical application of binary code clone detection to locate known vulnerable code in binary programs~\cite{pewny2015cross,eschweiler2016discovre,feng2016scalable}.
Given a piece of code which contains a known bug, it is possible to locate the corresponding clone~(or similar) code in other programs to check whether those programs are also vulnerable.

\texttt{NConvert}~\cite{nconvert} is a closed-source image processor which supports multiple formats.
It statically links the open-source library \texttt{libpng}~\cite{libpng} to handle files of the PNG format.
Function \emph{png\_set\_unknown\_chunks} of \texttt{libpng} is found to contain an integer overflow vulnerability before the version of 1.5.14~(\mbox{CVE-2013-7353}).
It is necessary to locate the statically-linked function in \texttt{NConvert} to verify whether the function is vulnerable and ensure the program security.
Since the source code of \texttt{libpng} is available, it is reasonable to fulfill the target with the clone code detection technique.
However, only the executable of \texttt{NConvert} is accessible that its compilation configuration is unknown.
Even though executables are compiled from the same code base, different compilation configurations would lead to semantics-equivalent transformation~(\textbf{C1}), generating syntax- and structure-variant binary code of equal semantics.
Hence, methods relying on syntax or structural features~(e.g., control flow graph isomorphism) become ineffective.
Besides, it is challenging to not only locate \emph{png\_set\_unknown\_chunks} accurately, but also achieve high code coverage of \texttt{NConvert}~(\textbf{C2}).
The target function is statically-linked, mixing with the user-defined functions of \texttt{NConvert}.
Static methods of binary code clone detection could cover all functions in \texttt{NConvert} to find \emph{png\_set\_unknown\_chunks}. However, they leverage semantics-less features, generating inaccurate results.
In contrast, dynamic methods depend on semantic features which are extracted via code execution, while they merely focus on the executed code.
It even requires huge extra work for dynamic methods to generate test cases in order to cover the target function.
Unfortunately, code coverage is still an issue for dynamic analysis of binaries~\cite{kargen2015turning}.

\subsection{System Overview of \binCMP}
We propose \binCMP\ to perform binary function clone analysis.
Given a binary function~(the \emph{template}),
\binCMP\ finds its clone match in the target binary program, returning a list of functions~(the \emph{targets}) from the program, which is ranked basing on the semantic similarity.

Figure~\ref{fig:sys_overview} presents the work flow of \binCMP.
Given the template function which has been well analyzed or understood~(\emph{png\_set\_unknown\_chunks}), \binCMP\ instruments and executes it with test cases, capturing its semantic signature~(\S\ref{sec:method:semantic_signature}).
Meanwhile, runtime information is recorded during the execution as well~(\S\ref{sec:method:instrument_execution}).
Then, \binCMP\ migrates the runtime information to each target function of the target binary program~(\texttt{NConvert}).
It emulates the execution of the target function to extract the semantic signature~(\S\ref{sec:method:emulation}).
Afterward, \binCMP\ compares the signature of the template function to that of each target function and computes their similarity score~(\S\ref{sec:method:similar_comparison}).
Lastly, a list of target functions is generated, which is ranked by the similarity scores in descending order.

In summary, to overcome \textbf{C1}, \binCMP\ completely depends on semantic signatures to detect binary function clones.
Additionally, the signatures are captured in a hybrid manner, which addresses \textbf{C2}.
\binCMP\ firstly extracts the signature of the template function via executing its test cases.
We assume that the template function has been well studied that its test cases are available.
In above example, the vulnerability of \emph{png\_set\_unknown\_chunks} has been known, and its test cases could be found in the \texttt{libpng} project as well as from the vulnerability database.
Then, with the runtime information of the template function, \binCMP\ generates the signature of each target function of the binary program under analysis~(\texttt{NConvert}) via emulation. Therefore, \binCMP\ is able to cover all target functions to detect their clone matches with signatures of rich semantics.

\section{Methodology}
In this section, we firstly introduce the semantic signatures adopted by \binCMP, then discuss how it captures the signatures of binary functions and measures their similarity.

\subsection{Semantic Signatures}\label{sec:method:semantic_signature}
For each binary function, \binCMP\ captures behaviors during the execution or emulation as its signature. Given a specific input, the signature indicates how the function processes the input and generates the output, reflecting the semantics of that function.
The signature consists of following features:

\begin{itemize}

\item \emph{Read and Written Values:}
The feature consists of global~(or static) variable values read from or written to the memory during an (emulated) execution.
It contains the input and output values of the function when provided with a specific input, indicating the semantics of the function.

\item \emph{Comparison Operand Values:}
The feature is composed of values for comparison operations whose results decide the following control flow of an (emulated) execution.
It indicates the path of the function followed by an input to generate the output.
Thus, it is semantics related as well.

\item \emph{Invoked Standard Library Functions:}
Standard library functions provide fundamental operations for implementing user-defined functions~(e.g.,~\texttt{malloc}, \texttt{memcpy}).
The feature has been shown to be semantics-related and effective for code clone analysis~\cite{wang2009begavior,wang2012can}.
Therefore, it is adopted as complement to the semantic signature of \binCMP.

\end{itemize}

During the execution or emulation, \binCMP\ captures the sequence of above features, and considers the sequence as the signature of a binary function for latter similarity comparison.


\begin{algorithm}[t]
	\SetKwInput{Input}{Input}
	\SetKwInOut{Output}{Output}
	\DontPrintSemicolon
	
	\caption{Algorithm of Instrumentation}
	\label{algthm:instrumentation}
	
	\Input{Instruction under Analysis $\mathcal{I}$}
	\Output{Instruction after Instrumentation $\mathcal{I}_r$}
	
	\Al{Instrumentation ($\mathcal{I}$)}{
		
		$\mathcal{I}_r \leftarrow \mathcal{I}$\\
		\tcp{capture features for the signature}
		\If{$\mathcal{I}$ accesses global/static data}{
			$\mathcal{I}_r \leftarrow$ {\ttfamily\bfseries record\_data\_val} ($\mathcal{I}_r$)
		}
		\If{$\mathcal{I}$ performs comparison}{
			$\mathcal{I}_r \leftarrow$ {\ttfamily\bfseries record\_oprd\_val} ($\mathcal{I}_r$)
		}
		\If{$\mathcal{I}$ calls a standard library function}{
			$\mathcal{I}_r \leftarrow$ {\ttfamily\bfseries record\_libc\_name} ($\mathcal{I}_r$)
		}
		
		\tcp{record runtime information}
		\If{$\mathcal{I}$ reads an argument of the function}{
			$\mathcal{I}_r \leftarrow$ {\ttfamily\bfseries record\_arg\_val} ($\mathcal{I}_r$) 
		}
		\ElseIf{$\mathcal{I}$ calls a function indirectly}{
			$\mathcal{I}_r \leftarrow$ {\ttfamily\bfseries record\_func\_addr} ($\mathcal{I}_r$)
		}
		\ElseIf{a function returns}{
			$\mathcal{I}_r \leftarrow$ {\ttfamily\bfseries record\_ret\_val} ($\mathcal{I}_r$)
		}

		\KwRet{$\mathcal{I}_r$}
	}
\end{algorithm}

\subsection{Instrumentation and Execution}\label{sec:method:instrument_execution}
In this step, \binCMP\ instruments a binary function \emph{F} to generate its signature by running test cases.
Meanwhile, runtime information for Emulation~(\S\ref{sec:method:emulation}) is recorded as well.

Algorithm~\ref{algthm:instrumentation} presents the pseudo-code of instrumentation.
\binCMP\ traverses each instruction~($\mathcal{I}$) of \emph{F}.
If $\mathcal{I}$ accesses global variables, performs comparison operations, or calls a standard library function, \binCMP\ injects code before $\mathcal{I}$ to capture corresponding features and generate the signature of \emph{F}~(\mbox{Line 4-9}).

\mbox{Line 11-16} present code for recording runtime values of \emph{F}'s execution.
According to \emph{cdecl}, the default calling convention of IA-32 binaries, function arguments are prepared by callers and passed through the stack.
Therefore, if $\mathcal{I}$ reads a variable which is pushed onto the stack before the invocation of \emph{F}, \binCMP\ considers the variable as a function argument and records its value~(\mbox{Line 11-12}).
Besides, \binCMP\ records the addresses of subroutines invoked by \emph{F} indirectly~(\mbox{Line 13-14}).
The return values of all subroutines, including both user-defined functions and standard library functions, are recorded as well~(\mbox{Line 15-16}).


\begin{algorithm}[t]
	\SetKwInput{Input}{Input}
	\SetKwInOut{Output}{Output}
	\DontPrintSemicolon
	
	\caption{Algorithm of Emulation}
	\label{algthm:emulation}
	
	\Input{Emulated Memory Space of the Target Function $\mathcal{M}$}
	\Input{Runtime Value Set of the Template Function $\mathcal{S}$}
	
	\Al{Emulation ($\mathcal{M}$, $\mathcal{S}$)}{
		{\ttfamily\bfseries assign\_func\_arg} ($\mathcal{M}$, $\mathcal{S}$)\\
		\ForEach{instruction $I$ to be emulated}{
			\If{$I$ reads global variables}{
				$addr \leftarrow$ {\ttfamily\bfseries get\_var\_addr} ($I$)\\
				\If{$addr$ is accessed for the first time}{
					{\ttfamily\bfseries migrate\_var\_val} ($\mathcal{M}$, $addr$, $\mathcal{S}$)
				}
			}
			\If{$I$ calls a function indirectly}{
				$addr \leftarrow$ {\ttfamily\bfseries get\_tar\_addr} ($I$)\\
				\If{$addr \in \mathcal{S}$}{
					{\ttfamily\bfseries migrate\_ret\_val} ($\mathcal{M}$, $addr$, $\mathcal{S}$)
				}
				\lElse{
					{\ttfamily\bfseries exit\_emulation}()
				}
			}
			\If{$I$ invokes a standard library function}{
				$libc \leftarrow$ {\ttfamily\bfseries get\_func\_name} ($I$)\\
				\If{$libc$ needs system supports}{
					{\ttfamily\bfseries migrate\_ret\_val} ($\mathcal{M}$, $libc$, $\mathcal{S}$)
				}
			}
			\tcp{capture features for the signature}
			\If{$I$ contains features}{
				{\ttfamily\bfseries record\_feat\_val} ($\mathcal{M}$, $I$)
			}
			{\ttfamily\bfseries emulate\_inst} ($\mathcal{M}$, $I$, $\mathcal{S}$)
		}
	}
\end{algorithm}

\subsection{Emulation}\label{sec:method:emulation}
For every target function \emph{T} to be compared with the template function \emph{F}, \binCMP\ emulates its execution with the runtime information extracted from the last step.
The semantic signature of \emph{T} is captured simultaneously.
Clone functions should behave similarly if they are executed with the same input~\cite{egele2014blanket}.
Namely, if \emph{T} is the clone match of \emph{F}, their signatures should be similar.
Algorithm~\ref{algthm:emulation} presents the pseudo-code of emulation. 
\binCMP\ provides \emph{T} with the arguments of \emph{F}~(Line~2), and emulates it with the runtime information of \emph{F}~(Line~20).
Besides, \binCMP\ records the features of \emph{T} to generate its signature~(\mbox{Line 18-19}).
Next, we discuss the algorithm for emulation in more details.

\subsubsection{Function Argument Assignment}
In our scenario, binary functions for comparison are compiled from the same code base, i.e.,~clone functions have the same number of arguments.
According to the calling convention, \binCMP\ recognizes the arguments of the target function \emph{T}.
If the argument number of \emph{T} equals to that of \emph{F}, \binCMP\ assigns argument values of \emph{F} to those of \emph{T} in order.
Otherwise, \binCMP\ skips the emulation of \emph{T} which cannot be the match of \emph{F}.
For example, \emph{F} and \emph{T} have the following argument lists:
\begin{center}
	\ttfamily\small
	\emph{F}(farg\_0, farg\_1, farg\_2)\\
	\emph{T}(targ\_0, targ\_1, targ\_2)
\end{center}
If \binCMP\ has the values of \texttt{farg\_0} and \texttt{farg\_2} that \emph{F} only accesses the tow arguments in the execution, \binCMP\ assigns their values to \texttt{targ\_0}, \texttt{targ\_2} separately.
To make the emulation smoothly, arguments without corresponding values~(\texttt{targ\_1}) are assigned with a predefined value~(e.g.,~0xDEADBEEF).

\subsubsection{Global Variable Reading}

In the execution of the template function \emph{F}, it might read global~(or static) variables whose values have been modified by former executed code.
To emulate the target function \emph{T} in the memory space of \emph{F}, \binCMP\ migrates global variable values of \emph{F} to corresponding addresses which \emph{T} reads from.
\binCMP\ needs to consider two points:
\begin{enumerate*}[label=\roman*)]
	\item getting the global variable addresses which \emph{T} reads from~(Line~5 of Algorithm~\ref{algthm:emulation}), and
	\item migrating the corresponding global variable value from the memory of \emph{F} to that of \emph{T}~(Line~7 of Algorithm~\ref{algthm:emulation}).
\end{enumerate*}

Global variables are stored in specific sections of a binary program~(e.g.,~\texttt{.data}).
The size of each variable is decided by the source code.
The location of the variable, including the base address of a global data structure~(e.g.,~array), is determined in the binary code after compilation and not changed afterward.
Thus, global variables are accessed with hard-coding addresses.
Each member of a global data structure is accessed by adding its corresponding offsets to the constant base address, and the offset is generated from the input~(function arguments).
Hence, \binCMP\ is able to obtain global variable addresses of \emph{T} easily during the emulation.

\begin{figure}[t]
	\vspace{-14.75pt}
	\captionsetup[subfloat]{captionskip=0pt}
	\subfloat[Template Function~(\emph{F})\label{fig:global_template}]{\usebox{\globaltemplate}}
	\hfill
	\subfloat[Target Function~(\emph{T})\label{fig:global_target}]{\usebox{\globaltarget}}
	\caption{Global Variable Value Migration}
	\label{fig:global_migration}
\end{figure}

\binCMP\ migrates global variable values according to their usage order.
Figure~\ref{fig:global_migration} shows an example of two functions for global variable value migration.
During the execution of \emph{F}, two global variables \texttt{gvar1} and \texttt{gvar2} are read at Line~1 and Line~3 separately in Figure~\ref{fig:global_template}.
\texttt{gvar1} is used to test its value at Line~2, and \texttt{gvar2} is used for the addition operation at Line~4.
So the usage order of the two variable is \mbox{\texttt{[gvar1}, \texttt{gvar2]}}.
When emulating \emph{T} in Figure~\ref{fig:global_target}, \binCMP\ identifies \texttt{ecx} and \texttt{ebp} are loaded with global variables \texttt{gvar1'} and \texttt{gvar2'} at Line~1 and Line~2.
Then, it finds \texttt{ebp} is used for testing at Line~3, and \texttt{ecx} is used for the addition at Line~4 afterward.
The usage order of the global variables in Figure~\ref{fig:global_target} is \mbox{\texttt{[gvar2'}, \texttt{gvar1']}}.
Therefore, \binCMP\ assigns the value of \texttt{gvar1} to \texttt{gvar2'}, and \texttt{gvar2} to \texttt{gvar1'} accordingly.
If there are no enough global values to assign~(e.g.,~\emph{T} reads two global variables but \emph{F} reads only one), \binCMP\ provides the surplus variables of \emph{T} with predefined values~(e.g.,~0xDEADBEEF).

\begin{figure}
	\centering
	\usebox{\switchjmp}
	\caption{Indirect Jump of a Switch}
	\label{fig:swt_jmp}
\end{figure}

\subsubsection{Indirect Calling/Jumping}
Targets of indirect calls are decided by the input at runtime.
Since the target function \emph{T} is emulated in the memory space of the template function \emph{F},
if \emph{T} is the clone match of \emph{F},
the indirect call targets of \emph{T} should be those invoked during the execution of \emph{F}.
\binCMP\ then migrates the return values of \emph{F} to corresponding indirect calls of \emph{T}~(\mbox{Line 10-11} in Algorithm~\ref{algthm:emulation}).
Otherwise, the target function under emulation cannot be the match of \emph{F}. \binCMP\ stops the process and exits~(Line~12 in Algorithm~\ref{algthm:emulation}).

An indirect jump~(or branch) is implemented with a jump table which contains an ordered list of target addresses.
Jump tables are stored in \texttt{.rodata}, the read-only data section of an executable.
Therefore, similar to the reading of a global data structure, a jump table entry is accessed by adding the offset to the base address of the jump table.
The base address is a constant value, and the offset is computed from the input.

Figure~\ref{fig:swt_jmp} shows an indirect jump of a switch structure.
At Line~2, the index value is computed with \texttt{edx}, a value of an input-related local variable, and stored in \texttt{eax}.
If the index value is not above \texttt{0x2A}, which represents the default case, an indirect jump is performed according to the jump table whose base address is 0x808F630~(Line~5).
As entries of a jump table are sorted, with identical input, clone code would have equal offset and jumps to the path of the same semantics.
\binCMP\ just follows the emulation and has no need to do extra work for indirect jumps.

\subsubsection{Standard Library Function Invocation}\label{sec:method:emulation:lib_func_emulation}
If the target function \emph{T} calls a standard library function which requests the system support~(e.g.,~\texttt{malloc}), \binCMP\ skips its emulation and assigns it with the result of the corresponding one invoked by the template function \emph{F}~(\mbox{Line 15-16}).
For example, \emph{F} and \emph{T} calls following library functions in sequence:
\begin{center}
	\ttfamily\small
	F: malloc\_0, memcpy, malloc\_1\\
	T: malloc\_0', memset, malloc\_1'
\end{center}
\binCMP\ assigns return values of \texttt{malloc\_0}, \texttt{malloc\_1} to \texttt{malloc\_0'}, \texttt{malloc\_1'} separately, and skips the emulation.
\texttt{memset} is emulated normally, because it has no need for the system support.

\subsection{Similarity Comparison}\label{sec:method:similar_comparison}
\binCMP\ has captured the semantic signature~(feature sequence) of the template function via execution, and those of target functions via emulation.
In this step, it computes the similarity score of the template function signature and that of each target function in pairs.
We utilize the Longest Common Subsequence~(LCS) algorithm~\cite{bergroth2000survey} to the similarity measurement.
On one hand,
a signature is captured from the~(emulated) execution of a function.
The appearance order of each entry in the signature is a feature as well.
On the other hand,
a signature is captured from optimized or obfuscated binary programs that it contains diverse or noisy entries in the sequence.
LCS not only considers the element order of two sequences for comparison,
but also allows skipping non-matching elements, which tolerates code optimization and obfuscation.
Hence, the LCS algorithm is suitable for signature similarity comparison of \binCMP.

The similarity score is measured by the Jaccard Index~\cite{hamers1989similarity}.
Given two semantic signatures $S_f$ and $S_t$, the Jaccard Index is calculated as followed:
\begin{equation}
J(S_f, S_t) = \frac{|S_f \cap S_t|}{|S_f \cup S_t|} = \frac{|S_f \cap S_t|}{|S_f| + |S_t| - |S_f \cap S_t|}
\end{equation}
Here, $|S_f|$ and $|S_t|$ are the lengths of sequence $S_f$ and $S_t$. $|S_f \cap S_t|$ is the LCS length of the two sequences.
$J(S_f, S_t)$ ranges from $0$ to $1$, which is closer to $1$ when $S_f$ and $S_t$ are considered more similar.

After this step, \binCMP\ generates a target function list which is ranked by the similarity scores in descending order.

\section{Implementation}\label{sec:implemetation}
Currently, \binCMP\ supports IA-32 binary function clone analysis of ELF~(Executable and Linkable Format) files.
Next, we discuss the key aspects of the implementation.

\subsection{Binary Function Boundary Identification}
\binCMP\ requires addresses and lengths of binary functions to perform clone analysis.
Given an ELF file, We leverage \mbox{\texttt{IDA Pro} v6.6}~\cite{idapro2006}, an industrial strength reverse engineering tool, to disassemble it, identifying the boundaries of each binary function.
The plugin of \mbox{IDA Pro}, \texttt{IDAPython}, provides interfaces to obtain addresses of functions, i.e.,~\mbox{\texttt{Functions(start, end)}} which returns a list of function first addresses between \texttt{start} and \texttt{end}.
Therefore, we develop a script with IDAPython to acquire function addresses of binary files automatically.
Although the resulting disassembly of \mbox{IDA Pro} is not perfect~\cite{andriesse2016an}, it is sufficient for \binCMP.

\subsection{Instrumentation and Emulation}\label{sec:imple:instr_emul}
We implement the instrumentation module of \binCMP\ with \texttt{Valgrind}~\cite{nethercote2007valgrind}, a dynamic instrumentation framework.
Valgrind unifies binary code under analysis into \texttt{VEX-IR}, a RISC-like intermediate representation~(IR), and injects instrumentation code into the IR code.
Then, it translates the instrumented IR code into binaries for execution.
IR translation unifies the operations of binary code and facilitates the process of signature extraction.
For example, memory reading and writing operations are all unified with \texttt{Load} and \texttt{Store}, the opcodes defined by VEX-IR.
Hence, we just concentrate on the specific operations of IR and ignore the complex instruction set of \mbox{IA-32}.

The step of emulation is implemented basing on \texttt{angr}~\cite{shoshitaishvili2016sok}, a static binary analysis framework.
angr borrows VEX-IR from Valgrind, translating binary code to be analyzed into IR statically.
Given a user-defined initial state, it provides a module named \texttt{SimProcedure} to emulate the execution of IR code.
SimProcedure allows injecting extra code to monitor the emulation of the IR code.
It actually emulates the process of instrumentation.
Besides, angr maintains a database of standard library functions to ease the emulation of those functions~(\S\ref{sec:method:emulation:lib_func_emulation}).
Thus, we develop a script of monitoring code, which is similar to the instrumentation code developed with Valgrind, to capture semantic signatures during the emulation with angr.

\subsection{Similarity Comparison}
As the length of a signature sequence might have the scale over $10^4$, the possibility is high for the traditional LCS algorithm, whose memory complexity is $O(mn)$, to run out the memory.
We implement \binCMP\ to compute LCS with the Hirschberg's Algorithm~\cite{hirschberg1975linear} which needs only $O(min(m, n))$ memory space.

\section{Evaluation}\label{sec:evaluation}
We conduct empirical experiments to evaluate the effectiveness and capacity of \binCMP.
Firstly, \binCMP\ is evaluated with binaries compiled with different compilation configurations, including variant optimization options and compilers. The results are then compared to those of existing solutions~(\S\ref{sec:eva:cc}).
Secondly, we evaluate the effectiveness of \binCMP\ in handling obfuscation by comparing binary programs with their obfuscated versions~(\S\ref{sec:eva:obf}).
Lastly, with the motivating example of \texttt{NConvert} described in \S\ref{sec:mot_ovv:mot}, 
we show how \binCMP\ locates the statically-linked defective function of \texttt{NConvert}~(\S\ref{sec:eva:case}).

\subsection{Experiment Setup}
The evaluation is performed in the system Ubuntu 16.04 which is running on an Intel Core \mbox{i5-2320 @ 3GHz} CPU with 8G DDR3-RAM.

\begin{table}[t]
	\renewcommand{\arraystretch}{1.3}
	\centering
	\caption{Object Projects of Evaluation}
	\label{tab:obj}
	\begin{tabularx}{.47\textwidth}{|c|c|X|}
		\hline
		Program & Version & \multicolumn{1}{c|}{Description} \\
		\hline
		\hline
		convert & 6.9.2 & Command-line interface to the ImageMagick image editor/converter \\
		\hline
		curl & 7.39 & Command-line tool for transferring data using various protocols \\
		\hline
		ffmpeg & 2.7.2 & Program for transcoding multimedia files \\
		\hline
		gzip & 1.6 & Program for file compression and decompression with the DEFLATE algorithm \\
		\hline
		lua & 5.2.3 & Scripting parser for Lua, a lightweight, multi-paradigm programming language \\
		\hline
		mutt & 1.5.24 & Text-based email client for Unix-like systems \\
		\hline
		openssl & 1.0.1p & Toolkit implementing the TLS/SSL protocols and a cryptography library \\
		\hline
		wget & 1.15 & Program retrieving content from web servers via multiple protocols\\
		\hline
	\end{tabularx}
\end{table}

\subsubsection{Dataset}

We adopt programs of eight real-world projects as objects of the evaluation, as listed in Table~\ref{tab:obj}.
The object programs have various functionalities, such as data compression~(\texttt{gzip}), code parsing~(\texttt{lua}), email posting~(\texttt{mutt}), etc.
With those object programs, the effectiveness of \binCMP\ is shown to be not limited by the type of programs and functions under analysis.

In the first group of experiments~(\S\ref{sec:eva:cc}), the object programs are compiled with different compilers, i.e.,~\mbox{\texttt{gcc} v4.7} and \mbox{\texttt{clang} v3.8.0}, and variant optimization options, i.e.,~\texttt{-O3} and \texttt{-O0}.
In the second group of experiments~(\S\ref{sec:eva:obf}), we adopt \mbox{Obfuscator-LLVM v4.0.1}~(\texttt{OLLVM})~\cite{ieeespro2015-JunodRWM} to obfuscate the object programs for comparison.
\texttt{OLLVM} provides three widely used techniques for obfuscation.
We leverage the three techniques to obfuscate the object programs which are optimized with \texttt{-O3} and \texttt{-O0} respectively.
Therefore, we compile \emph{10}~($=2\cdot 2 + 3\cdot 2$) unique binary executables for each object program, overall \emph{80}~($=10\cdot 8$) for the evaluation.

For each experiment, we select two from the \emph{10} executables of an object program, i.e.,~$E_{tem}$~(the template executable) and $E_{tar}$~(the target executable).
\binCMP\ executes $E_{tem}$ with test cases obtained from its project, considering each executed function as a template function.
Then it compares every template function to all target functions of $E_{tar}$ in pairs to find the clone match.
On average, \emph{291} functions are triggered in an execution of $E_{tem}$, and $E_{tar}$ contains \emph{4,353} functions.
As a result, \binCMP\ totally performs over \emph{100 million} pairs of function comparison in all the experiments.

\subsubsection{Ground Truth}
All the \emph{80} executables are stripped that their debug and symbol information is discarded for the evaluation.
To verify the correctness of the experimental results, we compile \emph{extra unstripped copies} of the 80 executables, and establish the ground truth with their debug and symbol information.

For each template function, \binCMP\ generates a list of target functions ranked by the similarity scores in descending order~(as described in \S\ref{sec:method:similar_comparison}).
According to the ground truth, if the symbol name of the \emph{Top 1} function~(with the highest similarity score) in the resulting list is the same as that of the template function, the match is considered to be \emph{correct}.
Besides, we manually verify cases of function inline.
For example, function \emph{A} invokes \emph{B} in $E_{tem}$, while the corresponding function \emph{B'} is inlined into \emph{A'} becoming \emph{A'B'} in $E_{tar}$, and \emph{B'} disappears.
If \binCMP\ matches \emph{B} with \emph{A'B'}, we consider it \emph{correct} as well.

\subsubsection{Evaluation Metrics}\label{sec:eva:setup:eva_metrics}
Similar to previous work~\cite{egele2014blanket,hu2016cross},
we measure the performance of \binCMP\ with accuracy, the percentage of executed template functions whose correct matches are found. The formula is as followed:
\begin{equation}
Accuracy = \frac{|Correct\ Matches|}{|Template\ Functions|}
\end{equation}


\subsection{Accuracy across Compilation Configurations}\label{sec:eva:cc}

\begin{figure*}[t]
	\subfloat[Gcc -O3 vs -O0\label{fig:cross_opt:gcc}]{\includegraphics[scale=.47]{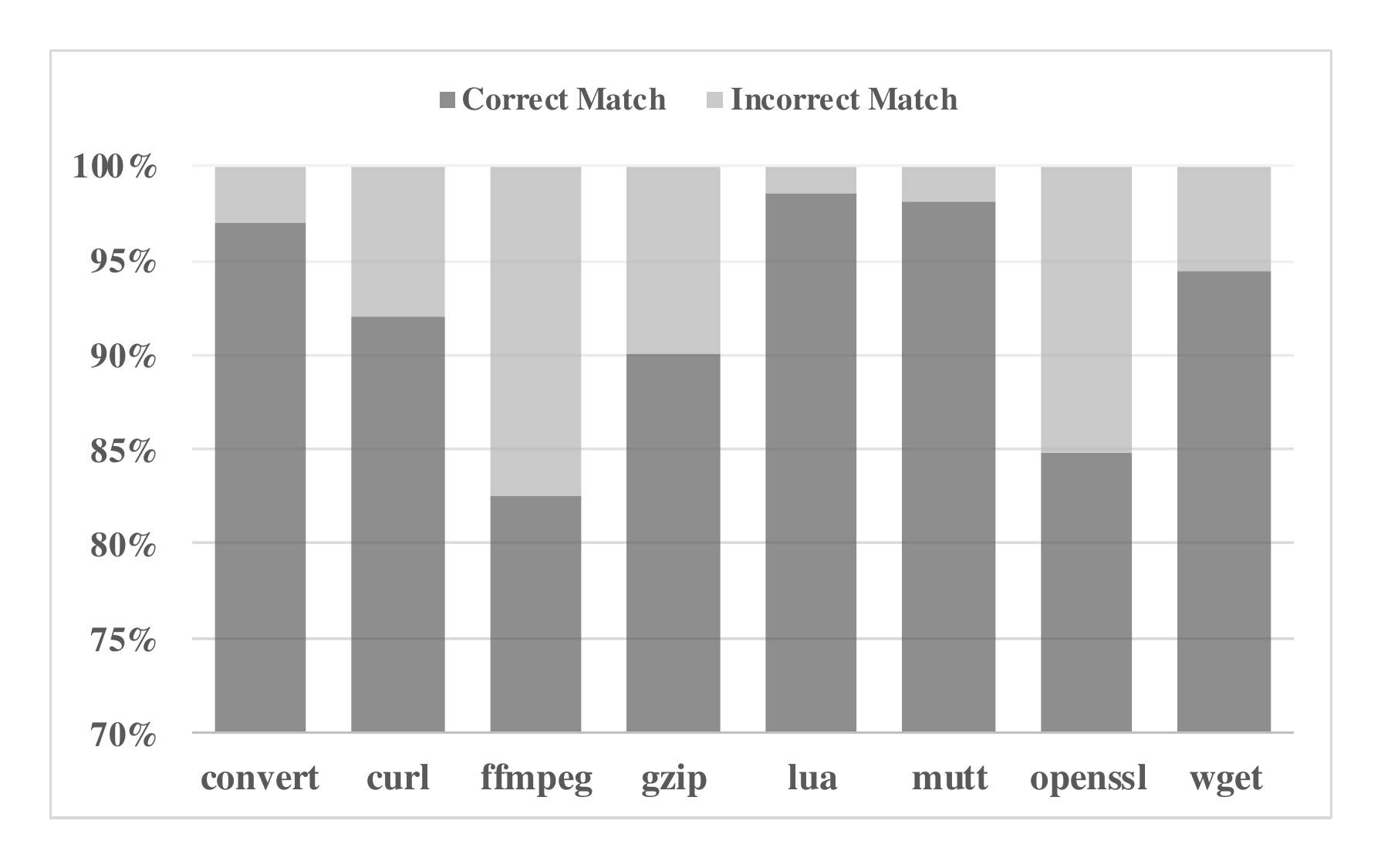}}
	\hfill
	\subfloat[Clang -O3 vs -O0\label{fig:cross_opt:clang}]{\includegraphics[scale=.47]{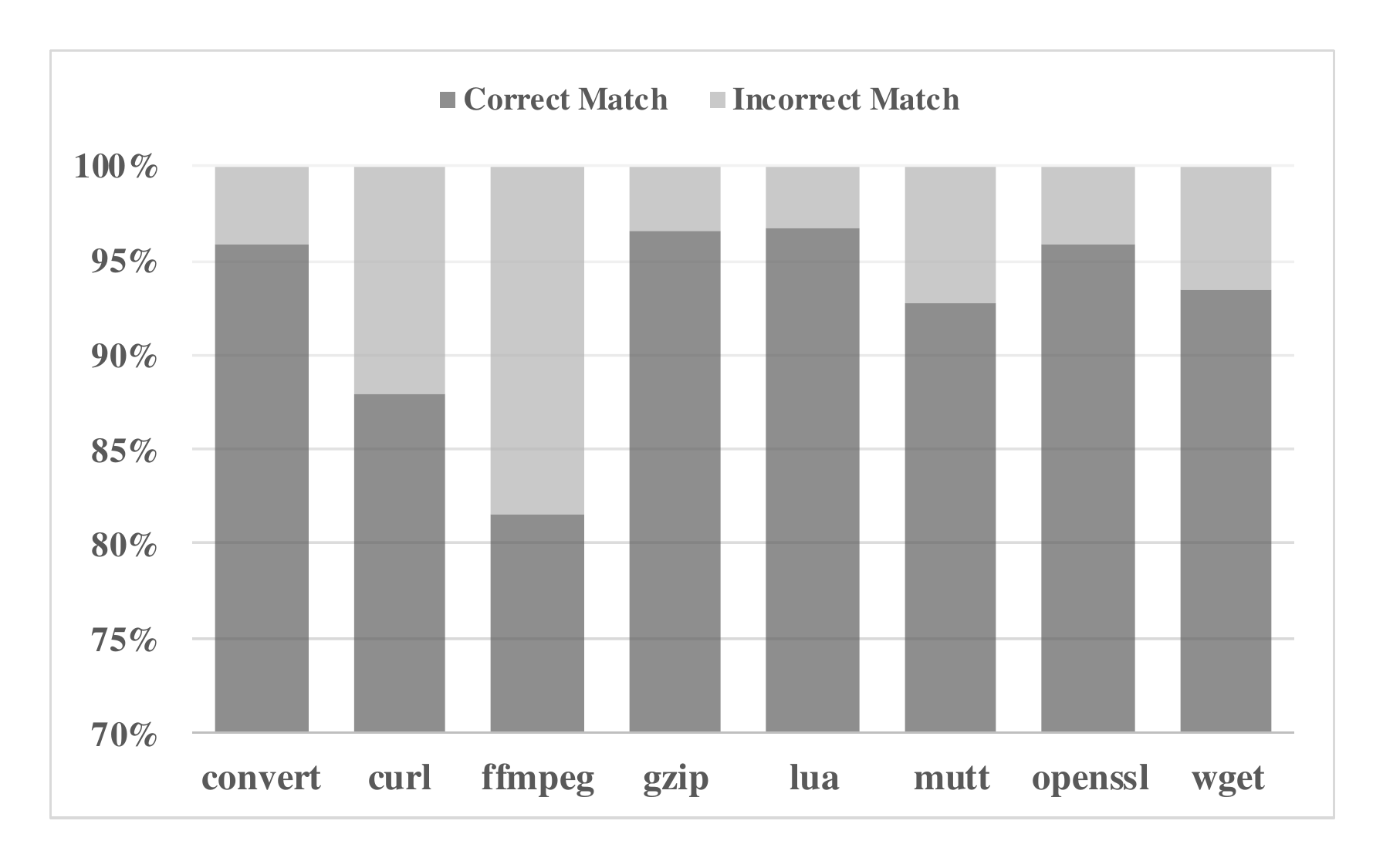}}
	\caption{Accuracy of Cross-optimization Analysis}
	\label{fig:cross_opt}
\end{figure*}

\subsubsection{Cross-optimization Analysis}\label{sec:eva:cc:opt}
In this section, we leverage \binCMP\ to match clone functions of different optimizations.
For a compiler, higher optimization options contain all strategies specified by lower ones.
Taking \texttt{gcc} as an example, the option \texttt{-O3} enables all \emph{88} optimizations of \texttt{-O2}, 
and turns on another \emph{14} optimization flags in addition.
Thus, we only discuss the case of \texttt{-O3}~($E_{tem}$) versus \texttt{-O0}~($E_{tar}$), which has larger differences than any other pair of cross-optimization analysis.

Figure~\ref{fig:cross_opt} shows the results of cross-optimization analysis for each object program compiled by \texttt{gcc}~(Figure~\ref{fig:cross_opt:gcc}) and \texttt{clang}~(Figure~\ref{fig:cross_opt:clang}) separately.
In Figure~\ref{fig:cross_opt:gcc}, \binCMP\ achieves the accuracy over \emph{82.0\%} for each object program, and the average accuracy is \emph{91.5\%}.
For every executable compiled by \texttt{clang} in Figure~\ref{fig:cross_opt:clang}, \binCMP\ correctly detects over \emph{80.0\%} functions of each object as well, and the average accuracy is \emph{92.0\%}.

We observe that \emph{function inline} is a reason leading to the incorrect matches.
For example, template \emph{A} calls \emph{B}, while the corresponding target function \emph{B'} is inlined into \emph{A'} becoming \emph{A'B'}.
Because the semantic signature of  \emph{A'B'} contains those of both \emph{A'} and \emph{B'},
signature length of \emph{A} is shorter than that of \emph{A'B'}.
Hence, the similarity score of function pair (\emph{A}, \emph{A'B'}) might be relative small, and \binCMP\ reports an incorrect match.

\begin{figure*}
	\subfloat[Gcc -O3 vs Clang -O0\label{fig:cross_com:gcc_O3_clang_O0}]{\includegraphics[scale=.47]{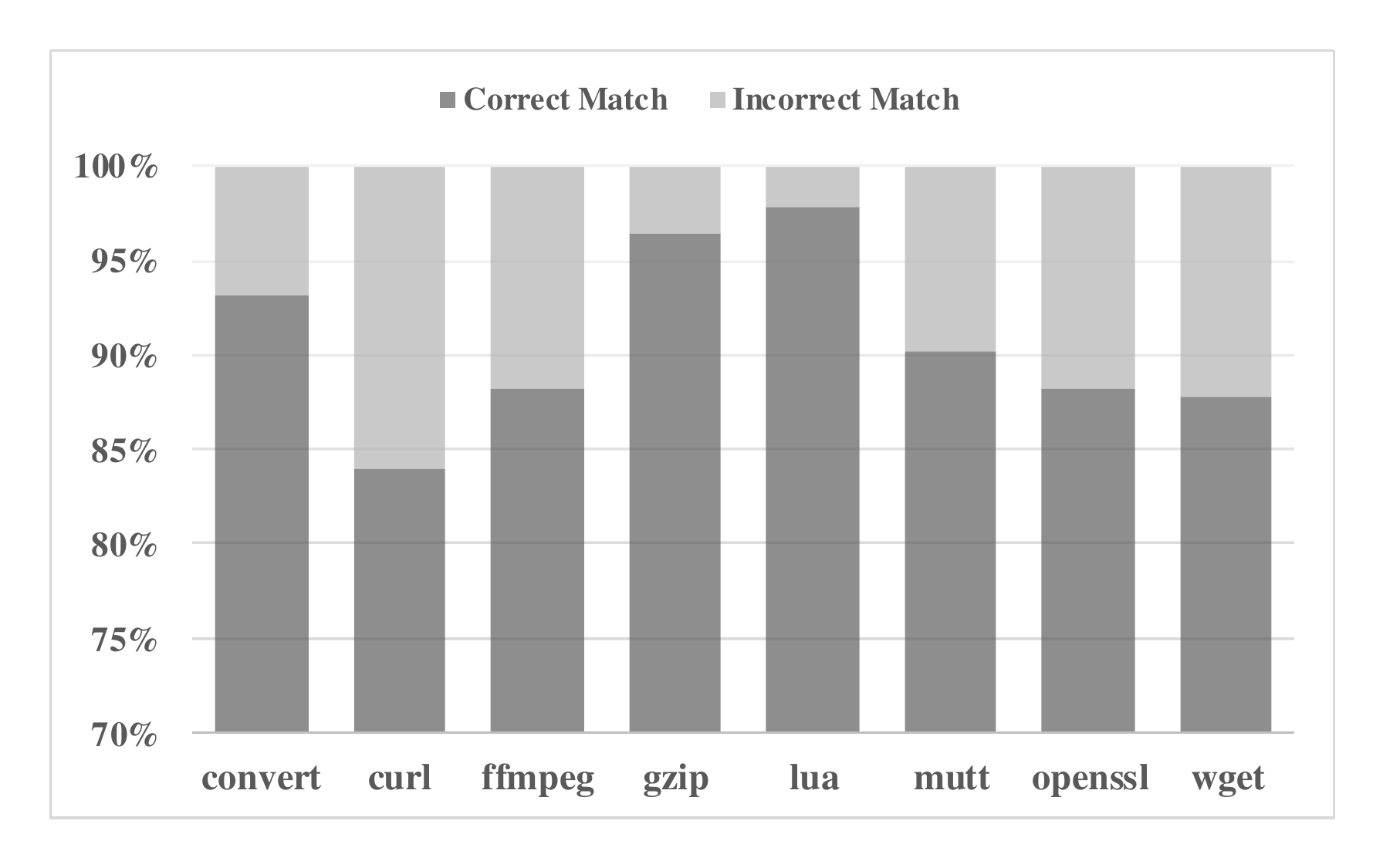}}
	\hfill
	\subfloat[Clang -O3 vs Gcc -O0\label{fig:cross_com:clang_O3_gcc_O0}]{\includegraphics[scale=.47]{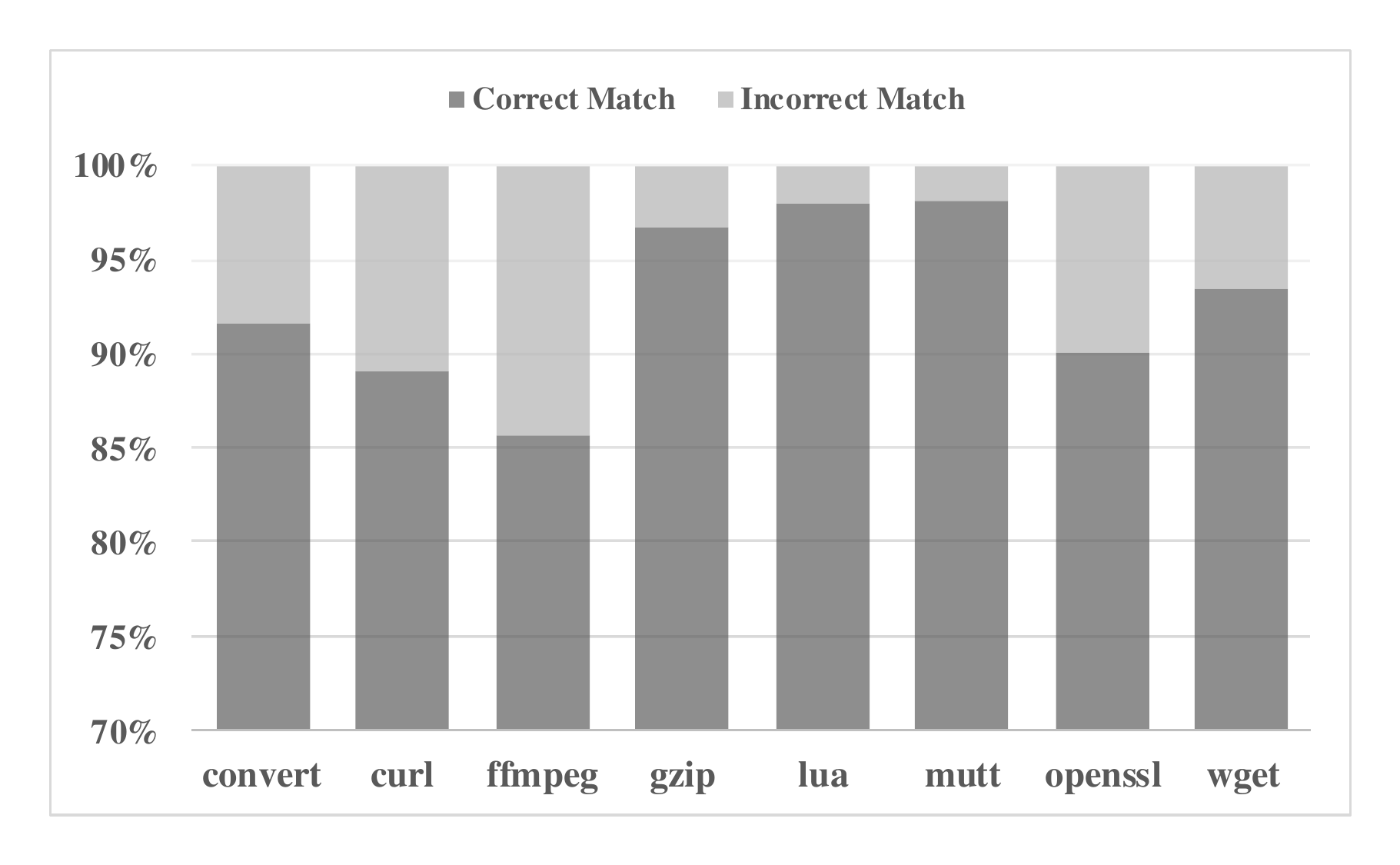}}
	\caption{Accuracy of Cross-compiler Analysis}
	\label{fig:cross_com}
\end{figure*}

\subsubsection{Cross-compiler Analysis}\label{sec:eva:cc:com}
In this section, \binCMP\ is evaluated with binaries compiled by different compilers.
Similar to the cross-optimization analysis, only the case of \texttt{-O3} versus \texttt{-O0} is considered.
The results are presented in Figure~\ref{fig:cross_com}.
For comparisons between \texttt{gcc -O3}~($E_{tem}$) and \mbox{\texttt{clang -O0}}~($E_{tar}$), 
\binCMP\ gives the accuracy all over \emph{84.0\%},
and the average accuracy is \emph{90.3\%}, as shown in Figure~\ref{fig:cross_com:gcc_O3_clang_O0}.
Additionally, in Figure~\ref{fig:cross_com:clang_O3_gcc_O0}, \binCMP\ achieves an average accuracy of \emph{91.4\%} for the setting of \mbox{\texttt{clang -O3}}~($E_{tem}$) versus \texttt{gcc -O0}~($E_{tar}$).
The accuracy of each object program exceeds \emph{85.0\%}.

In addition to function inline introduced by different optimizations, we find \emph{floating-point number} is another reason resulting in incorrect matches.
\texttt{gcc} leverages x87 floating-point instructions to implement corresponding operations, while \texttt{clang} uses the SSE~(Streaming SIMD Extensions) instruction set.
x87 adopts the FPU~(floating point unit) stack to assist in processing floating-point numbers. The operations deciding whether the stack is full or empty add redundant entries to the semantic signature of comparison operand values~(\S\ref{sec:method:semantic_signature}).
In contrast, SSE directly operates on a specific register set~(e.g.,~XMM registers) and has no extra operations.
Besides, x87 could handle single precision, double precision, and even 80-bit double-extended precision floating-point calculation, while SSE mainly processes single-precision data.
Due to the different precision of representations,
even though the floating-point numbers are the same, their values generated by the two compilers are not equal, which therefore results in the incorrect matches.

\begin{table}[t]
	\renewcommand{\arraystretch}{1.3}
	\centering
	\caption{Accuracy compared with the state-of-the-art solution Kam1n0 and the industrial standard tool BinDiff}
	\label{tab:tool_cmp}
	\begin{tabular}{|c|c|c|c|}
		\hline
		Setting & \textbf{\binCMP} & Kam1n0 & BinDiff \\
		\hline
		\hline
		gcc -O3 vs. gcc -O0 & 0.915 & 0.294 & 0.331 \\
		clang -O3 vs. clang -O0 & 0.920 & 0.252 & 0.506 \\
		\hline
		gcc -O3 vs. clang -O0 & 0.903 & 0.216 & 0.385 \\
		clang -O3 vs. gcc -O0 & 0.914 & 0.273 & 0.501 \\
		\hline
		\hline
		Average Accuracy & 0.916 & 0.265 & 0.409 \\
		\hline
	\end{tabular}
\end{table}

\subsubsection{Comparison with Existing Work}\label{sec:eva:cc:art}
In this section, we compare \binCMP\ to the state-of-the-art solution Kam1n0~\cite{ding2016kam1n0} and the industrial standard tool \mbox{BinDiff v4.2.0} supported by Google~\cite{flake2004structural,bindiff} from the perspective of detection accuracy.
Because Kam1n0 and BinDiff are both made available to the public, we could use the two solutions to detect binary clone functions with the same settings as \binCMP.
\binCMP\ is evaluated by the detection accuracy of executed template functions.
To perform fair comparison, we measure the accuracy of Kam1n0 and BinDiff by detecting clones of those template functions as well.
The results are presented in Table~\ref{tab:tool_cmp}.
Obviously, \binCMP\ outperforms Kam1n0 and BinDiff in detecting binary clone functions across compilation configurations.

Kam1n0 and BinDiff are typical solutions which rely on syntax and structure features to detecting binary clone functions.
Kam1n0 captures features of a function from its control flow graph~(CFG), and encode the features as a vector for indexing. Thus, essentially, it detects clone functions by analyzing graph isomorphism of CFG.
The relatively low accuracy of Kam1n0 indicates that compilation configurations indeed affect representations of binaries, even though two pieces of code are compiled from the same code base.
In addition to measuring the similarity of CFG, BinDiff considers other features to detect clone functions, such as function hashing which compares the hash of raw function bytes, call graph edges which match functions basing on the dependencies in the call graphs, etc. 
By carefully choosing suitable features to measure the similarity of functions, BinDiff becomes resilient towards variant compilers as well as optimization options to an extent.
Therefore, it performs better than Kam1n0, but is still at an apparent disadvantage compared to  \binCMP.

\subsubsection{Processing Time}
\binCMP\ analyzes binary function clones in three steps: instrumentation \& execution, emulation, and similarity comparison~(as shown in Figure~\ref{fig:sys_overview}).
According to Algorithm~\ref{algthm:instrumentation}, \binCMP\ injects code only to record runtime information and semantic signatures, and does not do online analysis.
Thus, the overhead of instrumentation is low.

As described in Emulation~(\S\ref{sec:method:emulation}), \binCMP\ merely emulates the user code of a target function, and borrows runtime values directly from the template function, skipping the emulation of specific system operations, such as allocating memory, reading or writing a file, etc.
As a result, it would not take much time to emulate a function.
In above experiments, \binCMP\ spends \emph{4.3} CPU seconds emulating a function on average.
Since \binCMP\ adopts LCS to compute similarity scores whose time complexity is relative high, the step of similarity comparison occupies the most processing time.
In the experiments, it costs \emph{573.9} CPU seconds on average to complete a pair of function comparison.

\begin{table}
	\centering
	\caption{Accuracy of analyzing obfuscated code. The template binaries are compiled with \texttt{gcc -O3}. \texttt{OLLVM} adopts \texttt{clang} as its compiler.}
	\label{tab:obf}
	\begin{tabular}{|c|c|c|c|}
		\hline
		Target & Obfuscation & \textbf{\binCMP} & BinDiff \\
		\hline
		\hline
		ollvm -O3 & \multirow{2}{*}{\tabincell{c}{Instructions Substitution}} & 0.891 & 0.676 \\
		ollvm -O0 & & 0.887 & 0.302 \\
		\hline
		ollvm -O3 & \multirow{2}{*}{\tabincell{c}{Bogus Control Flow}} & 0.843 & 0.295 \\
		ollvm -O0 & & 0.796 & 0.281 \\
		\hline
		ollvm -O3 & \multirow{2}{*}{\tabincell{c}{Control Flow Flattening}} & 0.874 & 0.464 \\
		ollvm -O0 & & 0.791 & 0.323 \\
		\hline
		\hline
		\multirow{2}{*}{\tabincell{c}{Average\\Accuracy}} & \multirow{2}{*}{\tabincell{c}{/}} & \multirow{2}{*}{\tabincell{c}{0.847}} & \multirow{2}{*}{\tabincell{c}{0.411}} \\
		& & & \\
		\hline
	\end{tabular}
\end{table}

\subsection{Accuracy of Matching Obfuscated Code}\label{sec:eva:obf}
In this section, we conduct experiments to compare normal binary programs with their corresponding obfuscated code.
We compile the object programs with the setting of \mbox{\texttt{gcc -O3}} as the normal code~($E_{tem}$).
We use all the three obfuscation methods provided by \texttt{OLLVM} to obfuscate the object programs generated with \mbox{\texttt{clang -O3}} and \mbox{\texttt{clang -O0}} separately~($E_{tar}$, \texttt{OLLVM} adopts \texttt{clang} as its compiler).

The experimental results are shown in Table~\ref{tab:obf}. Results of BinDiff are also presented as references.
\emph{Instruction substitution} replaces standard operators~(e.g.,~addition operators) with sequences of functionality-equivalent, but more complex instructions.
It obfuscates code on the \emph{syntax} level, affecting the detection accuracy of BinDiff, but posing few threats to \binCMP\ which is semantics-based.

\emph{Bogus control flow}~(BCF) adds opaque predicates to a basic block, which breaks the original basic block into two.
\emph{Control flow flattening}~(FLA) generally breaks a function up into basic blocks, then encapsulates the blocks with a selective structure~(e.g.,~the switch structure)~\cite{laszlo2009obfuscating}.
It creates a state variable for the selective structure to decide which block to execute next at runtime via conditional comparisons.
BCF and FLA both changes the \emph{structure} of the original function, i.e.,~modifying the control flow.
They insert extra code which is irrelevant to the functionality of the original function, generating redundant semantic features which are indistinguishable from normal ones~(e.g.,~comparison operand values of opaque predicates).
Thus, they affect the detection accuracy of \binCMP.
When analyzing binaries optimized with \texttt{-O0}, it correctly detects 79.6\% of functions obfuscated by BCF, and 79.1\% by FLA, while the average accuracy of \mbox{\texttt{gcc -O3} vs. \texttt{clang -O0}} is 90.3\%.
However, \binCMP\ still achieves twice the average accuracy of BinDiff, i.e.,~84.7\% of \binCMP\ and 41.1\% of BinDiff.

\subsection{Case Study: \texttt{libpng} vs. \texttt{NConvert}}\label{sec:eva:case}
As described in \S\ref{sec:mot_ovv:mot}, before the version of 1.5.14, \texttt{libpng} contains an integer overflow vulnerability in function \emph{png\_set\_unknown\_chunks}.
\texttt{NConvert}, a closed-source image processor, statically links the library to handle files of the PNG format.
In this section, we download the source code of \mbox{\texttt{libpng} v1.5.12} and the executable of \mbox{\texttt{NConvert} v6.17} from their home pages, aiming to locate \emph{png\_set\_unknown\_chunks} in \texttt{NConvert} with \binCMP.

We compile \mbox{\texttt{libpng} v1.5.12} with the default configurations, i.e.,~\mbox{\texttt{gcc -O2}}.
Then the clone analysis is performed in following steps:
\begin{enumerate}
	\item We run a test case of \texttt{libpng} from its project to cover the template function \emph{png\_set\_unknown\_chunks}, and recored its semantic signature as well as runtime information with \binCMP.
	As a result, the signature of the template function contains \emph{133} entries~(features).
	
	\item
	\emph{png\_set\_unknown\_chunks} has \emph{4} arguments.
	Thus, \binCMP\ only emulates \emph{337} target functions of \texttt{NConvert} whose identified argument number is \emph{4} as well.
	Overall, the process of emulation takes \emph{5987.0} CPU seconds.
	
	\item
	\binCMP\ totally spends \emph{57.0} CPU hours to compare the signature of the template function and those of \emph{337} target functions.
	It reports \emph{func\_81ad770}~(the function at 0x81AD770 in \texttt{NConvert}) achieves the highest similarity score \emph{0.378}~($=\frac{99}{133+228-99}$, the signature length is \emph{228} and LCS length is \emph{99}).
\end{enumerate}

By manual verification, we find the result is correct that \emph{func\_81ad770} is the clone match of \emph{png\_set\_unknown\_chunks}.
After locating the target function, analysts could do further analysis on it, e.g., checking whether the function is vulnerable, which is out of the scope of this paper.

\subsection{Threats to Validity}
\binCMP\ is implemented with \texttt{Valgrind} and \texttt{angr} which both adopt \texttt{VEX-IR} as the intermediate representation~(\S\ref{sec:imple:instr_emul}).
However, VEX-IR is not perfect that 16\% x86 instructions could not be lifted, although only a small subset of instructions is used in executables in practice and VEX-IR could handle most cases~\cite{kim2017testing}.
The incompleteness of \mbox{VEX-IR} might affect the accuracy of semantics signature extraction, while \binCMP\ still produces promising results in above experiments.

\section{Discussion and Future Work}
\subsection{Scope of Application}
In the evaluation, \binCMP\ is shown to be effective in analyzing obfuscated binary clone code which is generated by \texttt{OLLVM}~(\S\ref{sec:eva:obf}).
The robust of \binCMP\ is due to the nature of dynamic analysis and the adoption of semantic signatures.
However, that does not mean \binCMP\ could handle all kinds of obfuscations.
Besides, the \texttt{OLLVM} code actually affects the accuracy of \binCMP\ in the experiments.
When analyzing benign code, \binCMP\ achieves higher average accuracy which is 91.6\%,
while the ratio of obfuscated code is 84.7\%.
In the literature, deobfuscation has been well studied~\cite{udupa2005deobfuscation,yadegari2015symbolic,yadegari2015generic}. Therefore, if \binCMP\ fails to detect an obfuscated function, it is a better choice to deobfuscate it firstly, then perform further analysis.

In this paper, we present \binCMP\ to analysis binary programs of ELF~(Executable and Linkable Format) on the IA-32 architecture.
Because the method is semantics-based, \binCMP\ could be ported to other platforms, for example, PE~(Portable Executable) files on Windows.
Besides, \binCMP\ is implemented basing on \texttt{Valgrind} and \texttt{angr} which both support cross architecture analysis.
Hence, \binCMP\ is applicable for multiple architectures, such as x86-64, ARM, MIPS, etc.
We leave it as future work.

\subsection{Inline Function Detection}
As discussed in the section of Evaluation Metrics~(\S\ref{sec:eva:cc:opt}), function inline poses a threat to the accuracy of \binCMP.
Empirically, a compiler inlines a function because the function is short and invoked for numerous times.
Namely, size and invocation times might be features of an inline function.
Thus, it is possible to detect the inline functions with machine learning techniques.
If a function is considered as the potential inline function, we could combine it to its callers and capture the signature.
That is left as future work.

\subsection{Scalability}
The step of similarity comparison~(\S\ref{sec:method:similar_comparison}) is the performance bottleneck of \binCMP.
It calculates the similarity score of two signatures with the LCS algorithm whose time complexity is high.
However, the comparisons of function pairs are unrelated to each other.
The step could be implemented in parallel to reduce the total processing time.
Besides, MinHash~\cite{shrivastava2014defense} is a possible solution for the similarity comparison.
It calculates the Jaccard Index directly without computing the LCS of the two signatures.
However, MinHash treats the signature sequence as a set, discarding the order information of elements in the sequence, which is a potential semantic feature~(as discussed in \S\ref{sec:method:similar_comparison}).
Therefore, MinHash is a trade-off between accuracy and efficiency.

\section{Related Work}\label{sec:related_work}
Code clone~(or similarity) analysis is a classic topic of software engineering.
Due to the code reuse of software development, automatically identifying clone code becomes a common requirement of software maintenance~(e.g.,~bug detection).
The technique is also applied in other fields, e.g., malware analysis of security.
In the last twenty years, researchers have made much effort into source code clone analysis, typically including CCFinder~\cite{kamiya2002ccfinder}, DECKARD~\cite{jiang2007deckard}, CloneDR~\cite{baxter1998clone}, CP-Miner~\cite{li2006cp}, etc.
As the focus of this paper is clone analysis on binary code, which has its own challenges and scenarios,
we would not talk about work on source code in more details.
Next, we mainly discuss the related work on binary code clone analysis.

Syntax and structural features are widely adopted to detect binary clone code.
S{\ae}bj{\o}rnsen~et al.~\cite{saebjornsen2009detecting} detect binary clone code basing on opcode and operand types of instructions.
Hemel~et al.~\cite{hemel2011finding} treat binary code as text strings and measure similarity by data compression. The higher the compression rate is, the more similar the two pieces of binary code are.
Khoo~et al.~\cite{khoo2013rendezvous} leverage n-gram to compare the control flow graph~(CFG) of binary code.
David~et al.~\cite{david2014tracelet} measure the similarity of binaries with the edit distances of their CFGs.
BinDiff~\cite{flake2004structural} and Kam1n0~\cite{ding2016kam1n0} extract features from the CFG and call graphs to search binary clone functions.

As discussed earlier in this paper, the main challenge of binary code clone analysis is semantics-equivalent code transformation, such as link-time optimization, obfuscation, etc.
Because of the transformation, representations of binary code are altered tremendously, even though the code is compiled from the same code base.
Therefore, syntax- and structure-based methods become ineffective, and semantics-based methods prevail.
Jhi~et al.~\cite{jhi2011value} and Zhang~et al.~\cite{zhang2012first} leverage runtime invariants of binaries to detect software and algorithm plagiarism.
Ming~et al.~\cite{ming2017binsim} infer the lineage of malware by code clone analysis with the system call traces as the semantic signature.
However, those solutions require the execution of binary programs and cannot cover all target functions.
\mbox{Egele et al. \cite{egele2014blanket}} propose blanket execution to match binary functions with full code coverage which is achieved at the cost of detection accuracy.
Luo~et al.~\cite{luo2014semantics} and Zhang~et al.~\cite{zhang2014program} detect software plagiarism by symbolic execution.
Although their work is resilient to code transformation, symbolic execution is trapped in the performance of SMT/SAT solvers which cannot handle all cases, e.g.,~indirect calls.
David~et al.~\cite{david2016statistical} decompose the CFG of a binary function into small blocks, and measure the similarity of the small blocks basing on a statistical model.
However, the boundaries of CFG blocks would be changed by code transformation, affecting the accuracy of the method.

More recently, with the prevalence of IoT devices, binary code clone analysis is proposed to perform across architectures.
Multi-MH~\cite{pewny2015cross}, discovRE~\cite{eschweiler2016discovre}, Genius~\cite{feng2016scalable} are proposed to detect known vulnerabilities and bugs in multi-architecture binaries via code clone analysis.
BinGo~\cite{chandramohan2016bingo} and CACompare~\cite{hu2017binary} are proposed to analyze the similarity of binary code across architectures as well.
However, discovRE and Genius still depend heavily on the CFG of a binary function.
Multi-MH, BinGo and CACompare sample a binary function with random values to capture corresponding I/O values as the signature, while the random values are meaningless that they merely trigger limited behaviors of the function. Therefore, it is difficult for them to cover the core semantics of binary code.

To sum up, the topic of binary code clone analysis mainly focuses on two points:
\begin{enumerate*}[label=\roman*)]
	\item what signature to adopt, such as opcodes and operand types~(syntax), CFG~(structure) and system calls~(semantics);
	
	\item how to capture the signatures, such as statically disassembling, dynamically running and sampling, etc.
\end{enumerate*}
\binCMP\ leverage the combination of read and written values, comparison operand values, and invoked library functions as the signature which is able to better reveal the semantics of binary code.
Besides, it captures the signature via both execution and emulation, which not only ensures the richness of semantics, but also covers all target functions to be analyzed.

\section{Conclusion}
In this paper, we propose \binCMP, a hybrid approach, to detect binary clone functions.
\binCMP\ executes the template function with its test cases, and migrates the runtime information to target functions in order to emulate their executions.
During the execution and emulation, \binCMP\ captures semantic signatures of the functions for clone analysis.
The experimental results show that \binCMP\ is robust to semantics-equivalent code transformation, including different compilation configurations and commonly-used obfuscations.
Besides, we show that \binCMP\ performs better than the state-of-the-art solutions to binary code clone analysis, such as BinDiff and Kam1n0.


\section{Acknowledgments}
We would like to thank the anonymous reviewers for their insightful comments which greatly help to improve the manuscript.
This work is partially supported by the Key Program of National Natural Science Foundation of China~(Grant~No.U1636217), the National Key Research and Development Program of China~(Grant~No.2016YFB0801200), and a research grant from the Ant Financial Services Group.

\bibliographystyle{abbrv}
\bibliography{ref_icsme_18}

\end{document}